\documentclass[aps,prl,twocolumn,amssymb,showpacs,superscriptaddress]{revtex4}

\usepackage{graphicx}
\usepackage[english]{babel}
\usepackage{bm}    
\usepackage{subeqnarray}

\def\reff#1{(\ref{#1})}
\newcommand{\be}{\begin{equation}}
\newcommand{\ee}{\end{equation}}
\newcommand{\<}{\langle}
\renewcommand{\>}{\rangle}

\def\spose#1{\hbox to 0pt{#1\hss}}
\def\ltapprox{\mathrel{\spose{\lower 3pt\hbox{$\mathchar"218$}}
 \raise 2.0pt\hbox{$\mathchar"13C$}}}
\def\gtapprox{\mathrel{\spose{\lower 3pt\hbox{$\mathchar"218$}}
 \raise 2.0pt\hbox{$\mathchar"13E$}}}

\newcommand{\scrc}{{\cal C}}

\newcommand{\scrm}{{\cal M}}

\newcommand{\scro}{{\cal O}}

\def\R{{\mathbb R}}

\def\Prob{{\mathbb P}}   

\renewcommand{\emptyset}{\varnothing}

\begin{document}

\title{Some geometric critical exponents for percolation
 and the random-cluster model}

\author{Youjin Deng}
\affiliation{Hefei National Laboratory for Physical Sciences at Microscale
   and Department of Modern Physics,
   University of Science and Technology of China,
   Hefei, Anhui 230026, China}
\author{Wei Zhang}
\affiliation{Department of Physics, Jinan University,
      Guangzhou 510632, China}
\author{Timothy M.~Garoni}
\affiliation{ARC Centre of Excellence for Mathematics and Statistics of Complex Systems,
   Department of Mathematics and Statistics, University of Melbourne, VIC~3010, Australia}
\author{Alan D. Sokal}
\affiliation{Department of Physics, New York University,
      4 Washington Place, New York, NY 10003, USA}
\affiliation{Department of Mathematics,
      University College London, London WC1E 6BT, United Kingdom}
\author{Andrea Sportiello}
\affiliation{Dipartimento di Fisica and INFN, Universit\`a degli Studi
    di Milano, via Celoria 16, I-20133 Milano, Italy}

\date{April 20, 2009; revised January 12, 2010}

\begin{abstract}
We introduce several infinite families
of new critical exponents for the random-cluster model
and present scaling arguments relating them to the $k$-arm exponents.
We then present Monte Carlo simulations confirming these predictions.
These new exponents provide a convenient way to determine $k$-arm exponents
from Monte Carlo simulations.
An understanding of these exponents also leads to
a radically improved implementation of the Sweeny Monte Carlo algorithm.
In addition, our Monte Carlo data allow us to conjecture an exact
expression for the shortest-path fractal dimension $d_{\min}$
in two dimensions:
$d_{\rm min}\stackrel{?}{=} (g+2)(g+18)/(32g)$
where $g$ is the Coulomb-gas coupling,
related to the cluster fugacity $q$ via $q=2+2\cos(g\pi/2)$
with $2 \le g \le 4$.
\end{abstract}


\maketitle
\begin{figure}[t]
\vspace*{4mm}
  \begin{center}
\setlength{\unitlength}{28pt}
\hspace*{-4mm}  
    \begin{picture}(8,4)
\put(0,0){\includegraphics[scale=0.7]{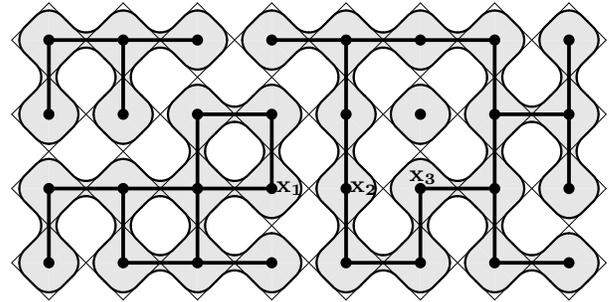}}
      \put(3.7,1.58){$\bf{x}_1$}
      \put(4.7,1.58){$\bf{x}_2$}
      \put(5.5,1.73){$\bf{x}_3$}
    \end{picture}
  \end{center}
\vspace*{-5mm}
\caption{
   Typical bond configuration $A\subseteq E$ on a finite subgraph
   of the square lattice:  here $|A|=31$ and $k(A)=4$.
   Also shown are the corresponding medial-lattice loops
   (interiors shaded for clarity).
   Here $C_{\min,2}=|\scrc_{\mathbf{x}_1}|=11$;
   choosing $e_1=\mathbf{x}_1\mathbf{x}_2$ and $e_2=\mathbf{x}_2\mathbf{x}_3$
   yields $M_{\min,2}=|\scrm_{\mathbf{x}_2,e_2}|=16$;
   and $T_{\min,2}=T_{\mathbf{x}_1}=4$.
   The graph-theoretic distance (shortest-path length)
   between $\mathbf{x}_2$ and $\mathbf{x}_3$ is 3.
  }

\label{diagram}
\end{figure}

 
The random-cluster model~\cite{RC_refs}
is a correlated bond-percolation model
that plays a central role in the theory of critical phenomena,
especially in two dimensions where it arises
in recent developments of conformal field theory \cite{DiFrancesco_97}
via its connection with Schramm--Loewner evolution (SLE) 
\cite{SLE_math,SLE_phys}.
To each bond configuration $A\subseteq E$ of a given graph $G=(V,E)$,
the random-cluster model assigns a weight proportional to $q^{k(A)}v^{|A|}$
where $k(A)$ is the number of clusters (connected components),
$|A|$ is the number of bonds, and $v,q\ge0$ are parameters
(see Fig.~\ref{diagram}).
For $q=1$
the random-cluster model
reduces to independent bond percolation \cite{percolation_refs},
while for integer $q \ge 2$ it provides a graphical representation
of the $q$-state ferromagnetic Potts model \cite{Potts_refs}.
Furthermore, the $q\to0$ limit corresponds to uniform spanning trees,
provided that $v\propto q^{\alpha}$ with $0<\alpha<1$;
this applies in particular to the
critical square-lattice random-cluster model,
for which $v=\sqrt{q}$ \cite{Baxter_book}.
The random-cluster model thus provides an extension of both percolation
and the Potts model that allows all positive values of $q$,
integer or noninteger, to be studied within a unified framework.

When the graph $G$ is planar,
it is useful to map the bond configurations to loop
configurations~\cite{BKW_76}, as illustrated in Fig.~\ref{diagram}.
The loop configurations are drawn on the medial graph, the vertices of
which correspond to the edges of the original graph.  The medial graph
of the square lattice is again a square lattice, rotated $45^\circ$.
Each unoccupied edge
of the original lattice is crossed by precisely two loop arcs, while
occupied edges are crossed by none. The continuum limits of such
loops are of central interest in studies of SLE \cite{SLE_phys}.
The outermost loop bounding a cluster defines its {\em hull}\/
\cite{Saleur-Duplantier}.

In this Rapid Communication we shall define and study several infinite families
of critical exponents for the random-cluster model, related to
cluster size, hull length and shortest path.
Some of these exponents have been studied previously for
spanning trees or percolation only, while others appear to be entirely new.
We shall give scaling arguments determining all but one of these exponents
as a function of $q$ in the two-dimensional (2D) case,
and Monte Carlo simulations confirming these predictions.
The remaining undetermined exponent is the
shortest-path fractal dimension $d_{\rm min}$~\cite{Stanley_84},
which relates the shortest-path length and Euclidean distance
between two vertices on a cluster.
However, our Monte Carlo data lead us to conjecture the exact formula 
\begin{equation}
d_{\rm min} \;\stackrel{?}{=}\; (g+2)(g+18)/(32g)
\label{dmin_conjecture}
\end{equation}
where $q=2+2\cos(g\pi/2)$ and
$g\in[2,4]$ is the Coulomb-gas coupling \cite{Nienhuis_84}.
To our knowledge, $d_{\min}$ has not previously been studied for $q \neq 0,1$.


Our original motivation for studying these exponents arose out of
practical concerns related to Monte Carlo algorithms, as we discuss below.
However, we subsequently discovered a relationship between them
and the $k$-arm exponents $x_k$ \cite{Saleur-Duplantier},
which have proved to be of fundamental importance in critical phenomena, 
not least within the context of rigorous studies of
percolation~\cite{rigorous_percolation}.
The exponents we study here are defined via the scaling
of very natural graphical observables, which are easily measured
in Monte Carlo simulations.
Therefore, the relationship between these exponents
and the $k$-arm exponents provides a new and convenient way
to numerically estimate $x_k$, which can be used
also in three dimensions (3D) where no exact expressions for $x_k$ are known.
This has allowed us, for example, to numerically verify in 3D that
$x_2$ equals the thermal dimension $x_t$ when $q=1$,
a result previously tested only in 2D.

\paragraph{Motivation.}
Monte Carlo simulations are an essential tool in statistical mechanics,
but they typically suffer from {\em critical slowing-down}\/
\cite{Sokal_Cargese_96}:
the autocorrelation (relaxation) time $\tau$ diverges
as a critical point is approached, most often as a power law
$\tau\sim\xi^{z}$, where $\xi$ is the spatial correlation length
and $z$ is a dynamic critical exponent.

The Sweeny algorithm~\cite{Sweeny_83} is a local single-bond update dynamics
for the random-cluster model;
for $0 < q < 1$ it is the {\em only}\/ known general algorithm for this model.
We have recently shown \cite{sweeny_prl} that
the Sweeny algorithm has unusually weak critical slowing-down,
fairly close to the theoretical lower bound
$z \ge \alpha/\nu$ \cite{Li-Sokal}.
Furthermore, it exhibits (especially for small $q$)
the surprising phenomenon of ``critical speeding-up'' \cite{sweeny_prl},
in which suitable global observables exhibit significant decorrelation
on time scales much {\em less}\/ than one sweep
(namely, $L^w$ hits for some $w < d$);
this makes the algorithm potentially very efficient.

The main obstacle to the use of the Sweeny algorithm
is the need (when $q \neq 1$) to determine how updating
a bond $\mathbf{x}\mathbf{y}$ affects $k(A)$:
this potentially requires traversing an entire cluster
to determine whether $\mathbf{x}$ and $\mathbf{y}$ are connected.
Given a lattice site ${\bf x}$,  we write $\scrc_{\bf x}$ for the
cluster containing ${\bf x}$ and $|\scrc_{\bf x}|$ for the number of
sites in it.
Since $|\scrc_{\bf{x}}|$ has mean $\sim L^{\gamma/\nu}$
near the critical point,
the connectivity check threatens to impose a
``computational critical slowing-down''
that would more than outweigh the good ``physical'' behavior
of the Sweeny dynamics. 
There do exist sophisticated algorithms
in the computer-science literature~\cite{dynamic_connectivity}
for performing such connectivity checking dynamically,
which have been proven to be (asymptotically) very efficient,
but their complexity appears prohibitive for use in practical simulations.

Our interest in the present project began with a simple idea for reducing
this computational slowing-down without the need for
complex data structures or algorithms:
namely, perform simultaneous breadth-first searches starting at
both endpoints $\mathbf{x}$ and $\mathbf{y}$,
and stop when one of the clusters has been fully visited
or the clusters merge. In the first case this takes a time
$\min(|\scrc_{\mathbf{x}}|,|\scrc_{\mathbf{y}}|)$,
and in the second case a time $B_{\rm s}$
(the number of sites visited in breadth-first search until merger).
A natural question is therefore to the determine
the critical behavior of $C_{\min,2}$ and $B_{\rm s}$,
where $C_{\min,2}=\min(|\scrc_{\mathbf{x}}|,|\scrc_{\mathbf{y}}|)$
if $\scrc_{\mathbf{x}}$ and $\scrc_{\mathbf{y}}$ are distinct
and $0$ otherwise.
We will provide here a scaling argument suggesting that
both scale (in mean) as $L^{d_F - x_2} \ll L^{\gamma/\nu}$,
where $d_F = d - \beta/\nu$ is the cluster fractal dimension
and $x_2$ is the 2-arm exponent.
We will then verify this prediction numerically.

In two dimensions an even more efficient procedure is to simultaneously
follow the medial-lattice loops surrounding $\mathbf{x}$ and $\mathbf{y}$.
If the two loops are distinct, this takes a time
$\min(|\scrm_{\mathbf{x},\mathbf{x}\mathbf{y}}|,
      |\scrm_{\mathbf{y},\mathbf{x}\mathbf{y}}|)$,
where $\scrm_{{\bf x},e}$ is the loop on the medial lattice
that winds around ${\bf x}$ through the edge $e$
if $e$ is unoccupied,
and $\scrm_{{\bf x},e} = \emptyset$ otherwise (see Fig.~\ref{diagram}).
This naturally leads to the question of the scaling of $M_{\min,2}$,
defined analogously to $C_{\min,2}$ as
$M_{\min,2}=\min(|\scrm_{\mathbf{x},\mathbf{xy}}|,
                 |\scrm_{\mathbf{y},\mathbf{xy}}|)$
if $\scrm_{\mathbf{x},\mathbf{xy}}$ and $\scrm_{\mathbf{y},\mathbf{xy}}$
are distinct and $0$ otherwise.
We will provide a scaling argument, and confirm numerically,
that $M_{\min,2}$ scales (in mean) as $\sim L^{d_H - x_2}$
where $d_H = 1 + 2/g$
is the hull fractal dimension~\cite{Saleur-Duplantier,note_loops}.
It follows that computational critical slowing-down is completely absent
for $q > 4 \cos^2(\pi \sqrt{2/3}) \approx 2.811520$.


\paragraph{Definition of exponents.}
These scaling results for $C_{\min,2}$ and $M_{\min,2}$
can be generalized in a very natural way.
We shall consider a variety of positive-integer-valued observables $\scro$; for each one we expect that its probability distribution obeys a scaling law
$\Prob(\scro=s) \sim s^{-\psi_\scro}$ (with $\psi_\scro > 1$) for large~$s$ at criticality in infinite volume,
or more generally
$\sim s^{-\psi_\scro} F_\scro(s/\xi^{d_\scro}, s/L^{d_\scro})$
near criticality in large finite volume, where $F_\scro$ is a scaling function.
Our goal is to determine, for each $\scro$,
the decay exponent $\psi_\scro$ and the fractal dimension $d_\scro$.
Note that
at criticality in finite volume,
$\< \scro^n \> \sim L^{(n+1-\psi_\scro) d_\scro} + {\rm const}$
+ corrections to scaling as $L \to\infty$
(or $\sim \log L$ if $n+1-\psi_\scro = 0$);
we shall use this fact 
in our Monte Carlo determinations of $\psi_\scro$ and $d_\scro$.


 
Fix nearby sites ${\bf x}_1,\ldots,{\bf x}_k$, and let
$C_{{\rm min},k} = \min(|\scrc_{{\bf x}_1}|,\ldots,|\scrc_{{\bf x}_k}|)$
if these clusters are all distinct, and 0 otherwise.
We expect that all these observables
have fractal dimension $d_{C_{{\rm min},k}}$
equal to the cluster fractal dimension $d_F = d - \beta/\nu$.
Moreover, standard hyperscaling arguments
\cite{percolation_refs}
give $\psi_{C_{{\rm min},1}} = d/d_F$
(the usual   
notation is $\tau = \psi_{C_{{\rm min},1}} + 1$).
Our goal is to determine $\psi_{C_{{\rm min},k}}$ for $k \ge 2$.
To our knowledge these exponents are new,
except $\psi_{C_{{\rm min},2}} = \frac{11}{8}$
for 2D spanning trees \cite{Manna_92}.

For two-dimensional lattices, choose for each site ${\bf x}_i$
a bond $e_i$ incident on it, and let
$M_{{\rm min},k} = \min(|\scrm_{{\bf x}_1,e_1}|,\ldots,|\scrm_{{\bf x}_k,e_k}|)$
if these loops are all distinct, and 0 otherwise. 
We expect that all these observables have fractal dimension
$d_{M_{{\rm min},k}}$ equal to the hull fractal dimension $d_H$;
and standard hyperscaling arguments give $\psi_{M_{{\rm min},1}} = d/d_H$.
We aim to determine $\psi_{M_{{\rm min},k}}$ for $k \ge 2$.

Now let $T_{\mathbf{x}}$ denote the maximum graph-theoretic distance
from $\mathbf{x}$ to any site in $\scrc_{\mathbf{x}}$,
and define $T_{{\rm min},k} = \min(T_{{\bf x}_1},\ldots,T_{{\bf x}_k})$
if the clusters $\scrc_{{\bf x}_i}$ are all distinct, and 0 otherwise
(see Fig.~\ref{diagram}).
We expect that all these observables have fractal dimension
$d_{T_{{\rm min},k}}$
equal to the shortest-path fractal dimension $d_{\rm min}$.
We aim to determine $\psi_{T_{\min,k}}$ for $k \ge 1$,
as well as $d_{\rm min}$.

Finally, consider a pair of sites separated by a distance $\sim L$,
say ${\bf x} = {\bf 0}$ and ${\bf x} = \boldsymbol{\alpha} L$
where $\boldsymbol{\alpha} \in \R^d$,
and let $S_{\boldsymbol{\alpha} L}$ be the length of the shortest path
connecting these sites if one exists, and 0 otherwise.
Since
$\Prob(\mathbf{0}\leftrightarrow \boldsymbol{\alpha}L)\sim L^{-2\beta/\nu}$,
we expect that
$\< (S_{\boldsymbol{\alpha} L})^n \> \sim L^{n d_{\rm min}- 2\beta/\nu}$.
We will use $\< (S_{\boldsymbol{\alpha} L})^n \>$
to numerically determine $d_{\rm min}$ in 2D,
leading to the conjecture (\ref{dmin_conjecture}).

\paragraph{Scaling arguments.}
Let $p_k(R)$ be the probability that, in an annulus of inner radius
$r \sim O(1)$ and outer radius $R$,
the inner circle is connected to the outer one by
at least $k$ distinct clusters. 
The $k$-arm exponent $x_k$ characterizes the large-$R$ asymptotics
of $p_k(R)$ at criticality:
$p_k(R) \sim R^{-x_k}$ as $R \to\infty$.
In particular, $x_1 = \beta/\nu$. 
In two dimensions it is known \cite{Nienhuis_84,Saleur-Duplantier,Cardy} that
\begin{subeqnarray}
   x_1  & = &  (g-2)(6-g)/(8g)  \\
   x_k  & = &  (g/8)k^2 - (g-4)^2/(8g)  \quad\hbox{for } k \ge 2 \quad
\end{subeqnarray}

Now fix nearby sites ${\bf x}_1,\ldots,{\bf x}_k$,
and let $C_k(s)$ be the probability that ${\bf x}_1,\ldots,{\bf x}_k$
belong to $k$ distinct clusters, each of which contains at least $s$ sites.
The correspondence $s \sim R^{d_F}$ suggests that
$C_k(s) \sim p_k(s^{1/d_F})$.
Since $C_k(s) = \Prob(C_{{\rm min},k} \ge s)$,
we predict $\psi_{C_{{\rm min},k}} = x_k/d_F + 1$
and hence $\< (C_{{\rm min},k})^n \> \sim L^{nd_F-x_k} + {\rm const}$.
This agrees with standard hyperscaling for $k=1$;
we will test it for $k \ge 2$. 
A similar argument suggests that $\< B_{\rm s}^n \> \sim L^{nd_F-x_2}$.
%



Analogous reasoning predicts $\psi_{M_{{\rm min},k}} = x_k/d_H + 1$
and $\psi_{T_{{\rm min},k}} = x_k/d_{\min} + 1$.
For $q=1$ this argument for shortest-path scaling is due to Ziff \cite{Ziff_99}.
In 2D it is known \cite{Saleur-Duplantier,Cardy}
that $d_H = 1+2/g$, but to our knowledge (\ref{dmin_conjecture}) 
is the only known conjecture for $d_{\min}$, even for $q=1$.

\paragraph{Monte Carlo simulations.}
We simulated the
two-dimensional
random-cluster model at criticality
for $q=0,0.01,0.25,0.5,1,1.5,2,3,3.5$ \cite{note_q=4_explanation}
and $4 \le L \le 1024$ (periodic boundary conditions)
by the Sweeny algorithm \cite{Sweeny_83} when $0 < q < 1$
and the Chayes--Machta algorithm \cite{Chayes-Machta} when $q > 1$.
For $0.25 \le q \le 2$ we also have data at $L=2048$.
For $q=1$ we used cluster-growth algorithms to handle $4 \le L \le 4096$.
For $q=0$ we used Wilson's algorithm \cite{Wilson_96}
to generate spanning trees from loop-erased random walk
\cite{note_q=0_explanation}.
The total CPU time used in these simulations was approximately 66 years using a 3.2 GHz Xeon EM64T processor.

For each observable $\scro$, we fit $\< \scro^n \>$ for $n=1,2,3$
to the Ans\"atze $a L^p$, $a L^p + b L^{p-\Delta}$ and $a L^p + c$,
varying $L_{\rm min}$ (the smallest $L$ value included in the fit)
until the $\chi^2$ was reasonable.  
The error bar is a subjective 68\% confidence limit that includes both statistical error and systematic error due to unincluded corrections to scaling.

\begin{figure}
\vspace*{-1mm}
\begin{center}
\includegraphics[scale=1]{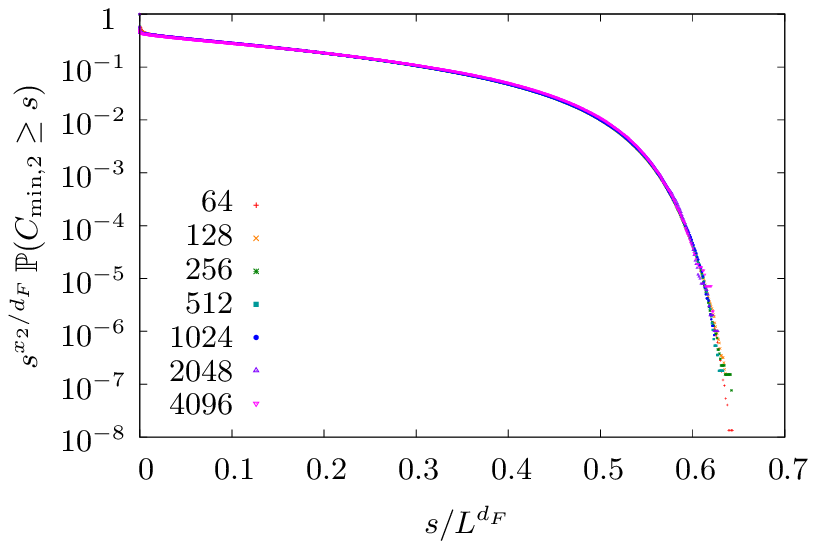}
\end{center}
\vspace*{-8mm}
\caption{
   (Color online)
   Finite-size-scaling plot showing
   $s^{x_2/d_F} \Prob(C_{{\rm min},2} \ge s)$
   versus $s/L^{d_F}$ for $q=1$ and $64 \le L \le 4096$.
}
\label{fig_FSS_Cmin2}
\end{figure}

Our results for $\< C_{{\rm min},2} \>$ and $\< M_{{\rm min},2} \>$
are presented in Table~\ref{table_Cmin};
complete results for $\< C_{{\rm min},2}^n \>$, $\< C_{{\rm min},3}^n \>$
and $\< M_{{\rm min},2}^n \>$ with $n=1,2,3$ can be found in~\cite{EPAPS}.
The agreement with the predicted exponents is excellent,
except where the exponent is very negative and hence
possibly overshadowed by correction-to-scaling terms.
A finite-size-scaling plot for $C_{{\rm min},2}$ and $q=1$
is shown in Fig.~\ref{fig_FSS_Cmin2} and exhibits excellent collapse.


\begin{table*}
\begin{center}
\footnotesize
\begin{tabular}{|c|c|lllllllll|}
\hline     
 Quantity  & \multicolumn{1}{c}{$q=$}
                  & \multicolumn{1}{c}{0}  & \multicolumn{1}{c}{0.01}
                  & \multicolumn{1}{c}{0.25}
                  & \multicolumn{1}{c}{0.5}  & \multicolumn{1}{c}{1}
                  & \multicolumn{1}{c}{1.5}  & \multicolumn{1}{c}{2}
                  & \multicolumn{1}{c}{3}  & \multicolumn{1}{c|}{3.5} \\
\hline
$\< C_{{\rm min},2} \>$ &
   $p$ (num.)     & 1.25002(4)   & 1.1800(1)     & 0.9273(2)     & 0.8071(2)     & 0.6458(1)
                  & 0.5226(2)    & 0.4163(3)     & 0.214(2)      & 0.108(3)  \\
 & $p$ (pred.)    & 1.25000      & 1.18007       & 0.92707       & 0.80768       & 0.64583
                  & 0.52298      & 0.41667       & 0.21667       & 0.10376 \\
\hline
$\< M_{{\rm min},2} \>$ &
   $p$ (num.)     & 1.25002(4)   & 1.1639(5)     & 0.8527(5)     & 0.7036(3)     & 0.4994(7)
                  & 0.3444(4)    & 0.2087(2)     & $-$0.0525(25)    & $-$0.195(3)  \\
 & $p$ (pred.)    & 1.25000      & 1.16439       & 0.85222       & 0.70341       & 0.50000
                  & 0.34420      & 0.20833       & $-$0.05000      & $-$0.19748   \\
\hline
$S_{L/2}$ &        $d_{\rm min}$ (num.)     &
  1.24999(3)   & 1.2371(10)    & 1.1825(3)  & 1.1596(4)  & 1.1303(8)   & 1.1112(7)    
& 1.0955(10)   & 1.0677(40)    & 1.0560(30)  \\
$2 \le g \le 4$  &   $d_{\rm min}$ (conj.)     &
 1.25000       & 1.23463       & 1.18211    & 1.15918    & 1.13021     & 1.10997    
&1.09375       & 1.06667       & 1.05343 \\
\hline
\end{tabular}
\caption{
   Numerical estimates versus theoretical predictions
   for exponents associated to $\< C_{{\rm min},2} \>$,
   $\< M_{{\rm min},2} \>$ and $S_{L/2}$.
}
\label{table_Cmin}
\end{center}
\end{table*}


%
%

Next we studied $S_{\boldsymbol{\alpha} L}$
for $\boldsymbol{\alpha} = (\frac{1}{2},0)$
in order to estimate $d_{\rm min}$:
see Table~\ref{table_Cmin} and \cite{EPAPS}.
Our result for $q=1$ is compatible with Grassberger's \cite{Grassberger_99}
estimate $d_{\rm min} = 1.1306(3)$.
The corrections to scaling are very strong
for these observables,
and our error bars are dominated by our assessment
of the likely systematic error from such corrections.
It would be very useful (but also very expensive)
to obtain data at larger values of $L$.

\begin{figure}
\vspace*{-2mm}
\begin{center}
\setlength{\unitlength}{50pt}
\begin{picture}(4.9,3.33)(0,0.1)
\put(0,0.1){\includegraphics[scale=0.68]{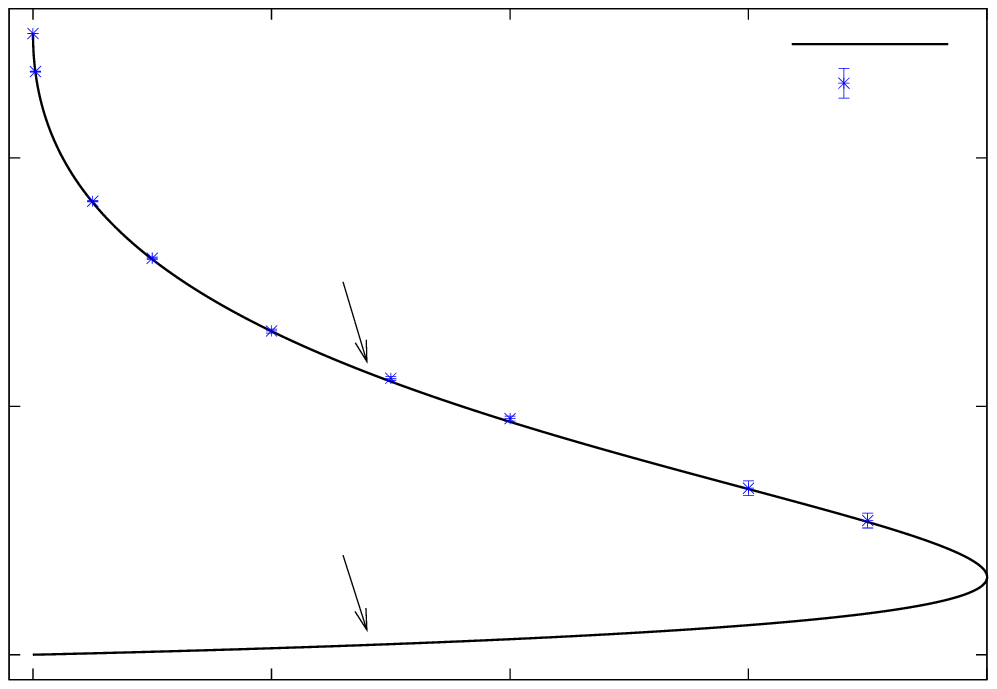}}
\put(0.8,0.33){$0$}
\put(1.74,0.33){$1$}
\put(2.68,0.33){$2$}
\put(3.62,0.33){$3$}
\put(4.56,0.33){$4$}
\put(4.36,0.26){$q$}
\put(0.4,0.6){$1.0$}
\put(0.4,1.58){$1.1$}
\put(0.4,2.56){$1.2$}
\put(0.25,3.){$d_{\rm min}$}
\put(3.3,2.17){\makebox[0pt][r]{critical ($2 \leq g \leq 4$)}}
\put(3.3,1.10){\makebox[0pt][r]{tricritical ($6 \geq g \geq 4$)}}
\put(3.75,3.01){\makebox[0pt][r]{predicted}}
\put(3.75,2.85){\makebox[0pt][r]{numerical}}
\end{picture}
\end{center}
\vspace*{-5mm}
\caption{
   (Color online)
   Numerical estimates for $d_{\rm min}$
   together with the conjectured exact formula
   $d_{\rm min} \stackrel{?}{=} (g+2)(g+18)/(32g)$.
}
\label{fig_dmin_q}
\end{figure}

Our results for $d_{\rm min}$ are consistent with the simple formula
\reff{dmin_conjecture}:
see Table~\ref{table_Cmin}
and Fig.~\ref{fig_dmin_q}.
This formula has the nice property that $d_{\rm min}$
is monotone decreasing for $2 \le g \le 6$
and reaches $d_{\rm min} = 1$ precisely at $g=6$,
in accordance with the idea \cite{Deng_backbone}
that clusters become more compact as $g$ grows.
It also agrees with the known fact
that $d_{\rm min} = \frac{5}{4}$ at $q=0$ \cite{Manna_92}.
Indeed, if one seeks a formula of the ``Coulomb-gas'' form
$d_{\rm min} = F(g) = Ag + B + C/g$ \cite{Nienhuis_84}
and imposes the constraints
$F(2) = \frac{5}{4}$, $F(6) = 1$, $F'(6) = 0$,
then the unique solution is \reff{dmin_conjecture}.

Our results for $T_{{\rm min},1}$ and $T_{{\rm min},2}$
are similar to those shown in Table~\ref{table_Cmin} and will be reported elsewhere.

\begin{acknowledgments}
We are grateful to Gary Reich for a discussion that inspired this work;
to Wenan Guo for participation in early stages of this project;
to Russ Lyons, Robin Pemantle and David Wilson for helpful correspondence;
and to NYU ITS for use of their computer cluster.
This research was supported in part by
NSF grant PHY--0424082,
NSFC grant 10975127,
and the Chinese Academy of Sciences.
\end{acknowledgments}





\begin{thebibliography}{99}

\bibitem{RC_refs}  P.W. Kasteleyn and C.M. Fortuin,
   J. Phys. Soc. Japan {\bf 26} (Suppl.), 11 (1969);
   C.M. Fortuin and P.W. Kasteleyn, Physica {\bf 57}, 536 (1972);
   G. Grimmett, {\em The Random-Cluster Model}\/
   (Springer, New York, 2006).

\bibitem{DiFrancesco_97}  P. Di Francesco, P. Mathieu and D. S\'en\'echal,
   {\em Conformal Field Theory}\/
   (Springer, New York, 1997).

\bibitem{SLE_math}  O. Schramm, Israel J. Math. {\bf 118}, 221 (2000);
   S. Rohde and O. Schramm, Ann. Math. {\bf 161}, 883 (2005);
   G.F. Lawler, {\em Conformally Invariant Processes
       in the Plane}\/ (American Mathematical Society, Providence, 2005).

\bibitem{SLE_phys} W. Kager and B. Nienhuis, J. Stat. Phys. {\bf 115},
      1149 (2004);
   J. Cardy, Ann. Phys. {\bf 318}, 81 (2005).

\bibitem{percolation_refs}  D. Stauffer and A. Aharony,
   {\em Introduction to Percolation Theory}\/, 2nd ed.
   (Taylor \& Francis, London, 1992);
   G. Grimmett, {\em Percolation}\/, 2nd ed.\ 
   (Springer, Berlin, 1999).

\bibitem{Potts_refs}  R.B. Potts,
   Proc. Cambridge Philos. Soc. {\bf 48}, 106 (1952);
   F.Y. Wu, Rev. Mod. Phys. {\bf 54}, 235 (1982);
   {\bf 55}, 315 (E) (1983);  J. Appl. Phys. {\bf 55}, 2421 (1984).

\bibitem{Baxter_book} R.J. Baxter, {\em Exactly Solved Models in Statistical
        Mechanics}\/ (Academic Press, London--New York, 1982).

\bibitem{BKW_76}  R.J. Baxter, S.B. Kelland and F.Y. Wu,
 J. Phys. A {\bf 9}, 397 (1976).

\bibitem{Saleur-Duplantier}  H. Saleur and B. Duplantier,
 Phys. Rev. Lett. {\bf 58}, 2325 (1987).

\bibitem{Stanley_84}  H. J. Herrmann, D. C. Hong and H. E. Stanley,
 J. Phys. A {\bf 17}, L261 (1984).

 \bibitem{Nienhuis_84}  B. Nienhuis, J. Stat. Phys. {\bf 34}, 731 (1984).

\bibitem{rigorous_percolation} M. Aizenman, B. Duplantier and A. Aharony,
     Phys. Rev. Lett. {\bf 83}, 1359 (1999);
  H. Kesten, Comm. Math. Phys. {\bf 109}, 109 (1987);
  S. Smirnov and W. Werner, Math. Res. Lett. {\bf 8}, 729 (2001).

\bibitem{Sokal_Cargese_96}
A.D. Sokal, in {\em Functional Integration: Basics and Applications}\/,
   ed. C. de Witt-Morette, P. Cartier and A. Folacci
   (Plenum, New York, 1997), pp.~131--192.

\bibitem{Sweeny_83}  M. Sweeny, Phys. Rev. B {\bf 27}, 4445 (1983).

\bibitem{sweeny_prl}  Y. Deng, T.M. Garoni and A.D. Sokal,
    Phys. Rev. Lett. {\bf 98}, 230602 (2007).

\bibitem{Li-Sokal}  X.-J. Li and A.D. Sokal,
   Phys. Rev. Lett. {\bf 63}, 827 (1989), see last paragraph.

\bibitem{dynamic_connectivity}
   M.R. Henzinger and V. King, J. ACM {\bf 46}, 502 (1999);
   J. Holm, K. de Lichtenberg and M. Thorup, J. ACM {\bf 48}, 723 (2001).

\bibitem{note_loops}
If the two loops are not distinct, we find that it takes (in mean) a 
time $\sim L^{d_H - x_2}$ to determine this also.
If the two loops are distinct and both topologically nontrivial,
connectedness needs to be determined by a full simultaneous cluster traversal.
Empirically this case arises with probability $\sim L^{-x_2}$
(and scaling arguments suggest that this is anyway an upper bound),
so that it makes a contribution
$\sim L^{d_F - 2x_2} \ll  L^{d_H - x_2}$ to the total cost.



\bibitem{Manna_92}  S.S. Manna, D. Dhar and S.N. Majumdar,
 Phys. Rev. A {\bf 46}, R4471 (1992).

\bibitem{Cardy} J. Cardy, J. Phys. A {\bf 31}, L105 (1998).













\bibitem{Ziff_99}  R.M. Ziff, J. Phys. A {\bf 32}, L457 (1999).

\bibitem{note_q=4_explanation}  We avoided $q=4$ because the
logarithmic corrections would likely make all exponent estimates unreliable
[see e.g.\ J. Salas and A.D. Sokal, J. Stat. Phys. {\bf 88}, 567 (1997)].

\bibitem{Chayes-Machta} L. Chayes and J. Machta,
   Physica A {\bf 254}, 477 (1998);
   Y. Deng {\em et al.}\/, Phys. Rev. Lett. {\bf 99}, 055701 (2007).

\bibitem{Wilson_96}  D.B. Wilson,
in {\em Proceedings of the Twenty-Eighth Annual ACM Symposium on the
  Theory of Computing}\/
  (ACM, New York, 1996), pp.~296--303.

\bibitem{note_q=0_explanation}
More precisely, we used Wilson's algorithm
to generate a random spanning tree $T$,
and then took measurements on $T \setminus T_0$,
where $T_0$ is a fixed short tree connecting ${\bf x}_1,\ldots,{\bf x}_k$.

\bibitem{EPAPS} See EPAPS Document No.~xxxxx for full tables.
For more information on EPAPS, see http://www.aip.org/pubservs/epaps.html.

\bibitem{Grassberger_99}  P. Grassberger, J. Phys. A {\bf 32}, 6233 (1999).



\bibitem{Deng_backbone}   Y. Deng, H.W.J. Bl\"ote and B. Nienhuis,
   Phys. Rev. E {\bf 69}, 026114 (2004).
Note that $4 \le g \le 6$ corresponds to the {\em tricritical}\/ Potts model
\cite{Nienhuis_84}.




%
%


%
%
%
%
%
%


%
%
%
%
%
%
%
%
%
%

%
%
%
%
%





\end{thebibliography}
\end{document}